\documentclass[letterpaper]{natureprintstyle}
\usepackage{graphicx,epsfig}
\usepackage{amsmath,amssymb,bbm}
\usepackage{braket}
\usepackage{eufrak}
\usepackage{color}
\usepackage[usenames,dvipsnames]{xcolor}
\usepackage[colorlinks=true,linkcolor=Red,citecolor=Green,linktoc=page]{hyperref}
\usepackage{multirow}

\usepackage[varg]{txfonts}
\usepackage{fancyhdr}

\usepackage[caption=false]{subfig}
\usepackage{balance}
\usepackage{float}
\usepackage{flushend}

\DeclareFontFamily{U}{rcjhbltx}{}
\DeclareFontShape{U}{rcjhbltx}{m}{n}{<->rcjhbltx}{}
\DeclareSymbolFont{hebrewletters}{U}{rcjhbltx}{m}{n}

\newcommand{\addresses}[1]{
\thispagestyle{fancy} 

\lfoot{\parbox{\textwidth}{ 
        \vspace{0.1cm}
        \rule{\textwidth}{0.2pt}
        \hspace{-0.2cm} \textsf{\scalefont{0.80} #1} \vspace{-0.2cm}
        \begin{center}{\scalefont{1} \thepage}\end{center}}}
\cfoot{} }

\newcommand{\onlinecite}[1]{\nocite{#1}\citenum{#1}}

\DeclareMathSymbol{\lamed}{\mathord}{hebrewletters}{108}

\begin{document}

% \title{Algorithmic quantum advantage %\textcolor{red}{reachability} 
% using a hybrid solution for linear systems of equations} 
\title{Large-scale quantum hybrid solution for linear systems of equations}

\author{M.\,R.\,Perelshtein,$^{1,2,3}$
        A.\,I.\,Pakhomchik,$^{1,2}$
        A.\,A.\,Melnikov,$^{1,2}$
        A.\,A.\,Novikov,$^{2}$
        A.\,Glatz,$^{4,5}$
        G.\,S.\,Paraoanu,$^{1,3}$
	 	V.\,M.\,Vinokur$^{1,6,\dagger}$
	 	and
	 	G.\,B.\,Lesovik$^{1,2}$
}

\date{\today}
\maketitle

\addresses{
    $^1$Terra Quantum AG, St.\,Gallerstrasse 16A, 9400 Rorschach, Switzerland.
	$^2$Moscow Institute of Physics and Technology, 141700, Institutskii
	Per.\ 9, Dolgoprudny, Moscow Distr., Russian Federation.
	$^3$QTF Centre of Excellence, Department of Applied Physics, Aalto University School of Science, P.O. Box 15100, FI-00076 AALTO, Finland.
	$^4$Materials Science Division, Argonne National Laboratory, 9700 S.
	Cass Avenue, Argonne, Illinois 60637, USA.  
    $^5$Department of Physics, Northern Illinois University,  DeKalb IL, 60115, USA.
    $^6$Physics Department, City College of the City University of New York, 160 Convent Ave, New York, NY 10031, USA.
	$^\dagger$e-mail: vv@terraquantum.swiss.
}

\date{}
\maketitle

\begin{abstract}
%A wealth of quantum algorithms developed during the past decades brought about the concept of quantum supremacy.
State-of-the-art noisy intermediate-scale quantum devices (NISQ), although
imperfect, enable computational tasks that are manifestly beyond the capabilities of modern classical supercomputers.
However, present quantum computations are restricted to exploring specific simplified protocols,
whereas implementation of full-scale quantum algorithms aimed at solving concrete large scale problems arising in data analysis and numerical modeling remains a challenge.
% Present quantum computations are severely restricted by the quantum processor power, 
Here we introduce and implement a hybrid quantum algorithm for solving linear systems of equations with exponential speedup, utilizing quantum phase estimation, one of the exemplary core protocols for quantum computing.
We introduce theoretically classes of linear systems that are suitable for current generation quantum machines and solve experimentally a $2^{17}$-dimensional problem on superconducting IBMQ devices, a record for linear system solution on quantum computers.
The considered large-scale algorithm shows superiority over conventional solutions, demonstrates advantages of quantum data processing via phase estimation, and holds high promise for meeting practically relevant challenges.
\end{abstract}

%\tableofcontents

\section*{Introduction}~~\\
\label{intro}
%\textcolor{red}{Наш contribution: 1) Приведен пример задачи, решение которой на квантовом компьютере сильно упрощается по сравнению с аналогами. Это может служить примером для ответа на вопрос: "Лучше решать задачу на квантовом или классическом компьютере?" (это я пытаюсь объяснить фразу These choices mean that the problem admits an efficient quantum circuit construction suitable for near-term quantum computers 2) Произведен анализ эффектинвость одна из важнейших квантовых алгоритмов на больших масштабах задачи.}
Since Peter Shor's discovery of the factorization algorithm in 1994\,\cite{Shor}, finding quantum algorithms for solving other classically intractable problems has become one of the mainstreams of computational science. 
The subsequent developments brought in the idea of an overwhelming advantage of quantum computing versus classical computing, often referred to as quantum supremacy\,\cite{Preskill2012}.
%After Peter Shor invented a quantum algorithm for integer factorization\cite{Shor}, promising to solve the problem exponentially faster than any classical factorization, a wealth of solutions based on quantum algorithms emerged, targeting classically intractable problems. 
%This sparked the idea of an overwhelming advantage of quantum computing, which in turn gave rise to the concept of quantum supremacy\cite{Preskill2012}. 
The manifestations of the latter sparked expectations for upcoming breakthrough provided quantum solutions would be implemented on noiseless quantum processor units (QPUs).
Yet the progress in quantum computing has been hindered by the inherent errors in logical gates, imperfections in readout procedures, and decoherence affecting qubits.
Despite these obstacles, state-of-the-art intermediate-scale quantum devices\,\cite{Preskill2018}, although suffering these noise-related problems, enable computations far beyond the capabilities of modern classical supercomputers.

While quantum advantage has recently been demonstrated on QPUs programmed to execute random instructions\,\cite{Supremacy}, the possibility of implementing specific transformations that can be adopted into a framework of practical quantum algorithms remains an open question.
% In Ref.\,\onlinecite{Supremacy} the Google collaboration demonstrated quantum supremacy by solving a problem that was intentionally designed to profoundly demonstrate the advantage of quantum over classical devices. 
% A low-noise superconducting 53-qubit QPU has been programmed to execute random instructions defined by a random quantum circuit containing single- and two-qubit gates.
% Such a random unitary transformation was designed to mimic possible gate-based quantum computations. 
% However, the possibility of implementing specific transformations that can be adopted into a framework of practical multi-qubit quantum algorithms remained an open question.
An exemplary core protocol for quantum computing is the quantum phase estimation\,\cite{nielsen} (QPE) protocol, which has been extensively studied and applied as a subroutine in a variety of quantum algorithms such as the Shor algorithm, quantum counting, and the calculation of the eigenvalues of unitary matrices.
By virtue of the QPE, a quantum computer gains an exponential speedup in the context of the matrix inversion problem\cite{Lloyd}.
%It could be benefitial to note at this point that there are algorithms that don't use QPE and in fact outperform those that do in terms of the precision parameter such as Childs, Kothari, Somma
% Это сюда вообще не вписывается, оно умаляет наше замечательное QPE.
The latter is a significant computational challenge in many areas, including artificial intelligence\,\cite{Lee2017}, partial differential equations\,\cite{Grossmann2007}, and data analysis\,\cite{freedman_2009}.

% In this paper, we demonstrate the superiority of quantum algorithms over classical ones by the example of one of the most important problems of solving systems of linear equations. 
% We were shown that the stimulation of a quantum circuit that solves the problem we set is exponentially faster than its classical counterparts. 
% In addition, we analyzed the effectiveness of solving this problem on a real quantum computer.

In our work, in order to exploit the QPE protocol for the fast matrix inversion, 
%\textcolor{red}{we introduce a linear system of equations that can be effectively, in the sense of minimum number of gates, solved on quantum computer using  hybrid version of the quantum Harrow--Hassidim--Lloyd algorithm.}
% Этому не место в Intro
we analyze and implement experimentally a hybrid version of the quantum Harrow--Hassidim--Lloyd algorithm\,\cite{Lloyd} (H-HHL) for solving linear systems of equations.
% This can serve as an example for answering the question: "Is it Better to solve a problem on a quantum or classical computer?" 
One of the main advantages of the HHL approach is the {\it exponential compression} of the data, which provides a notable improvement of the computational resource allocation.
%This allows a quantum algorithm to operate on only several dozen qubits to solve a large linear system, a task for which modern classical algorithms would fill the whole memory even of a high-end supercomputers.
Here, in order to most profoundly demonstrate such an advantage, we consider certain classes of linear systems that can be effectively solved with a quantum algorithm demonstrating an exponential speedup.
Our experiments, carried out on the high-end simulator and on IBMQ superconducting processors\,\cite{IBMQ}, demonstrate the superiority of quantum algorithms over the classical ones in solution of linear systems.
Besides, a thorough experimental analysis enables us to spot the shortcomings of the modern QPUs and outline solutions for surpassing them in order to boost a large-scale quantum data processing\,\cite{Biamonte2017}.

\begin{figure*}[t]
    \noindent\centering{
    \includegraphics[width=180mm]{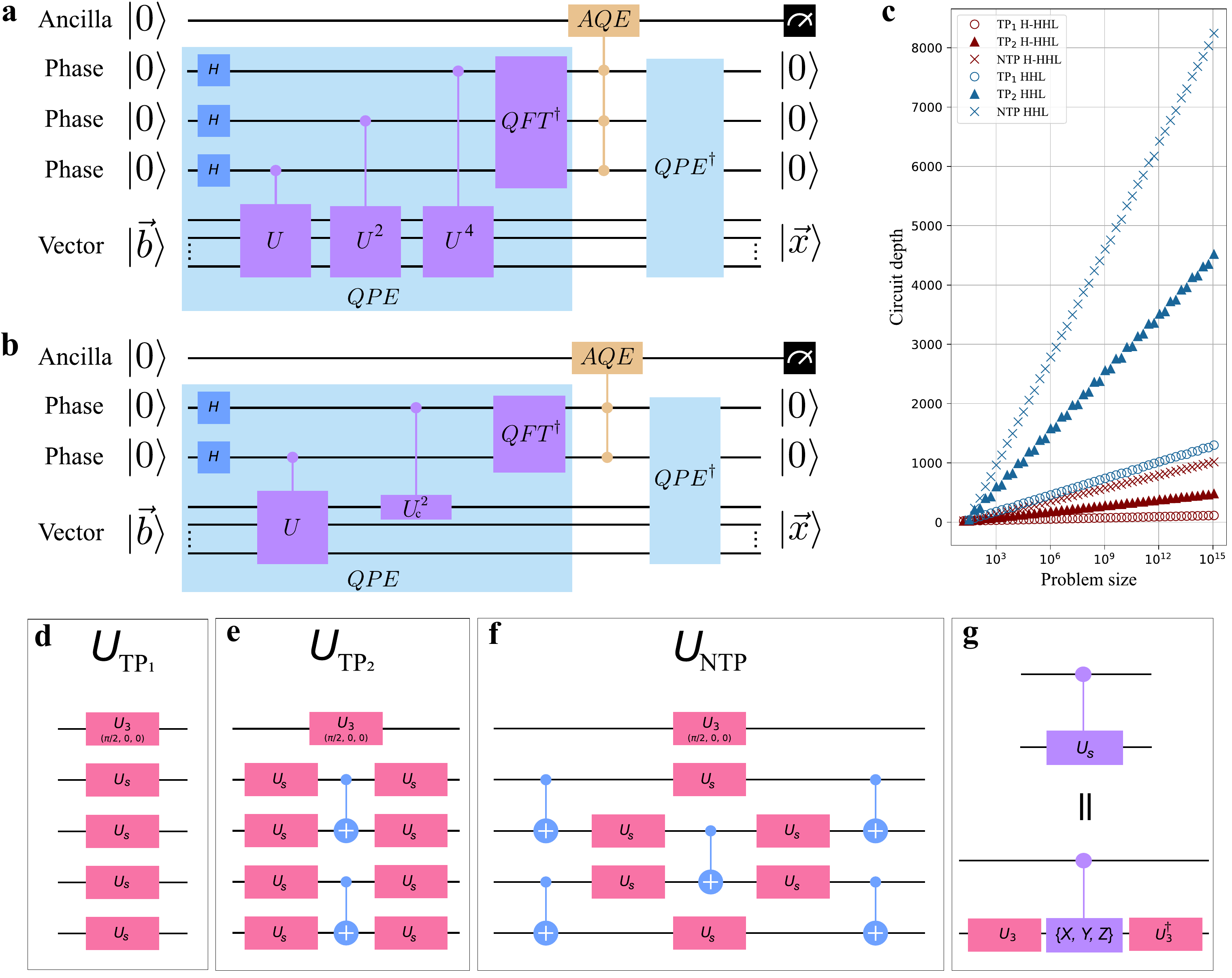}
    }
    \caption{
    {\bf a.}~Quantum scheme of the original HHL algorithm with 3 phase qubits that involves the quantum phase estimation protocol (QPE) and the ancilla quantum encoding step (AQE) in order to solve Eq.\,\ref{SLE} with a unitary matrix $\hat{U}$. 
    While the QPE part exploits controlled unitary operations $C\hat{U}$, $C\hat{U}^2$, $C\hat{U}^4$ and Quantum Fourier Transform in order to process eigenvalues and eigenvectors of $\hat{U}$, the AQE algorithm assists the matrix inversion.
    {\bf b.}~Quantum scheme of the simplified hybrid HHL algorithm, which provides the solution for the same matrix as the circuit from {\bf a}.
    The algorithm involves only 2 phase registers, and the QPE part is significantly reduced.
    {\bf c.}~The circuit depth as function of problem size for {\bf TP$_1$} (circles), {\bf TP$_2$} (triangles) and {\bf NTP} (crosses) matrix types processing by the original HHL and by the simplified H-HHL.
    {\bf d. e. f.} Quantum circuits that generate 3 types of $2^5\times2^5$ $\hat{U}$: $\hat{U}_{\bf TP_1}$ that involves only single-qubit gates $\hat{U}_s$; $\hat{U}_{\bf TP_2}$ that localizes two-qubit clusters with a single layer of CNOTs; $\hat{U}_{\bf NTP}$ that entangles all qubits expect the first.
    All types also contains $\hat{U}_3$ operator, which is used for the setting of the matrix spectrum.
    %Here, $\hat{U}_3(\pi/2,0,0)$, which we add to the first register in each type of $\hat{U}$, is a correcting gate that determines the matrix eigenvalues -- it was chosen in such way that the spectrum of any $\hat{U}$ is exp[$2\pi i\cdot\{\frac{1}{8}, \frac{3}{8}, \frac{5}{8}, \frac{7}{8}\}$].
    {\bf g.}~The scheme of a controlled-unitary gate $C\hat{U}_s$ of the single-qubit operation $\hat{U}_s$ that involves single two-qubit interaction.
    %Note that the $C\hat{U}_3$ implementation still involves two CNOTs.
    %Control-CNOT that is another component of $C\hat{U}$ is a Toffoli gate.
    %The $C\hat{U}_3(\pi/2,0,0)$ implementation involves two CNOTs, since we implement it using conventional technique\cite{nielsen}.
    %By constructing $C\hat{U}_s$, $C\hat{U}_3(\pi/2,0,0)$ and Toffoli gates one can easily build control-$\hat{U}$ operation, which is required in the QPE part.
    } 
    \label{quantum_circuit}
\end{figure*}

%We show that our implementation of the H-HHL algorithm offers a route to benchmarking quantum supremacy and to resolving the relevant practical challenges.

\bigskip

\section*{Results}

\subsection{Hybrid quantum algorithm}~~\\
%In this section we consider the hybrid quantum HHL algorithm and its implementation on modern quantum processors and discuss the main shortcomings associated with real superconducting QPUs.
% \textcolor{orange}{"When describing the HHL algorithm the quantum amplitude amplification step is omitted. That should be considered a part of the original approach. It should be emphasized that in this manuscript this step is omitted. Also worth discussing is the fact that this does not have a significant effect since the amplitude amplification step improves the complexity in terms of the condition number which for the problems considered in this work is O(1). "}
Solving a general system of linear equations $A\,\Vec{x}=\Vec{b}$ means finding the $N$-dimensional vector $x$ for a given $N\,\times\,N$ matrix $A$ and a vector $\Vec{b}$.
In this work we solve the linear system using the HHL quantum algorithm\,\cite{Lloyd}.

This algorithm employs matrix exponentiation $\hat{U}=e^{iA}$, or Hamiltonian simulation, as a preparation step. 
In general, such an exponentiation procedure is a major challenge, however for certain types of matrices this problem has been solved\,\cite{HamSimulation1,HamSimulation2}, with general approaches being presented in Ref.\,\onlinecite{TrotterSuzuki}.
Here, to focus on the phase estimation part of the quantum algorithm, we consider only particular matrices $A$ for which the Hamiltonian simulation can be efficiently performed.

Let us briefly discuss the structure of the quantum algorithm for linear systems. %that was firstly presented in Ref.\,\onlinecite{Lloyd}. 
The HHL algorithm that inverts a $N\,\times\,N$ matrix exploits three groups of qubits: the vector register that consists of $[\log{N}]$ qubits, the $p$-qubit phase register, and a single qubit ancilla register, $n = [\log{N}]+p+1$ qubits in total.
The whole computation, in turn, consists of the QPE algorithm, the ancilla quantum encoding (AQE) part, in which the single ancillary qubit conditionally operates on the state of the phase registers, and the inverse QPE.
Here, we omit the quantum amplitude amplification step\,\cite{Lloyd} that improves the complexity in terms of the conditional number, which is fixed.
%in the considered problem.
The phase estimation protocol exploits controlled unitary operations $C\hat{U}, C\hat{U}^2, \dots, $ $C\hat{U}^{2^{p - 1}}$ over phase and vector qubits and the Quantum Fourier Transform afterward in order to process eigenvalues and eigenvectors of the matrix.
For instance, the quantum circuit of the HHL algorithm that utilizes 3 phase registers is depicted on Fig.\,\ref{quantum_circuit}a.

In order to implement the HHL algorithm we use the spectral decomposition of the $N\,\times\,N$ matrix
\begin{equation}
    \hat{U}=\sum_{j=1}^{N} e^{2\pi i \cdot \lambda_j} \ket{u_j}\bra{u_j}
\end{equation} 
and encode the vector $\Vec{b}$ into the qubit's amplitudes $\ket{b} = \sum_{i=0}^{N}\,\beta_i\,\ket{i}$.
Once we run the algorithm, the solution is encoded into the quantum state
\begin{equation}
    \ket{x} = \frac{(\log{\hat{U}})^{-1}\ket{b}}{\mathcal{N}_x} = \frac{1}{\mathcal{N}_x}\sum_{j=1}^{N} \frac{1}{\lambda_j} \ket{u_j},
\end{equation}
where $\mathcal{N}_x$ is the normalization coefficient. 
Using projective measurement, we obtain the expectation value $\bra{x}\hat{M}\ket{x}$ for some operator $\hat{M}$ within exponentially shorter time than that allowed by the classical algorithms.
\begin{figure*}[t]
    \noindent\centering{
    \includegraphics[width=175mm]{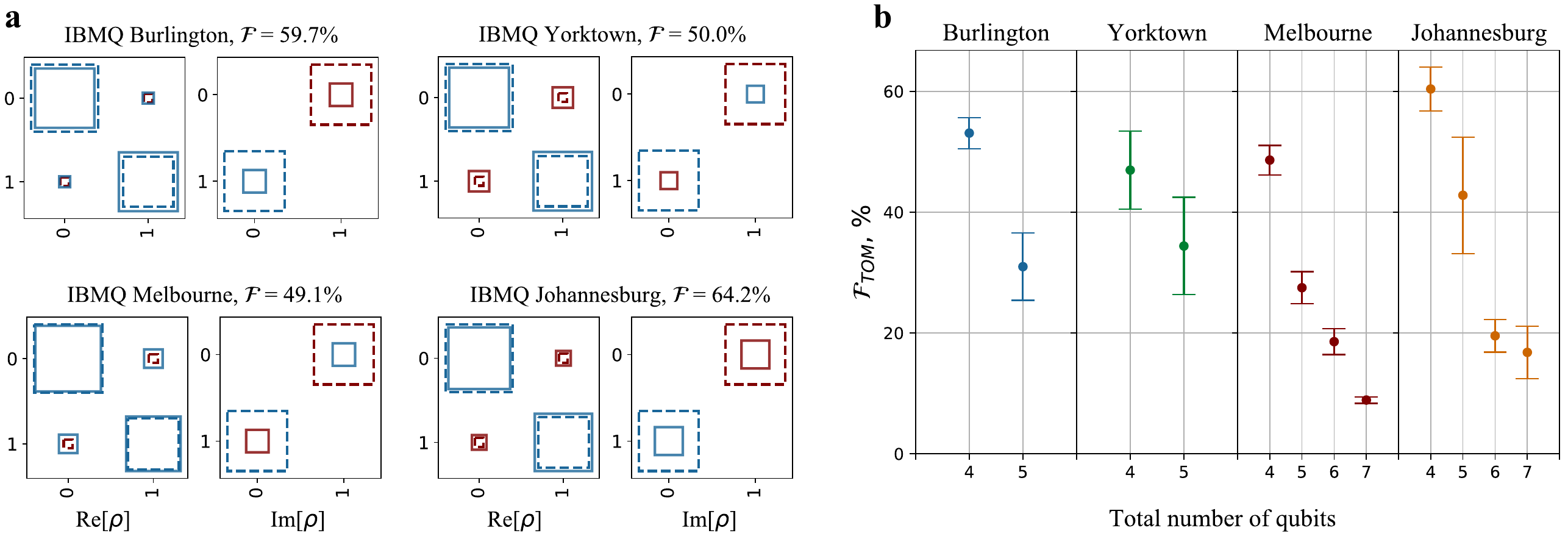}
    }
    \caption{
        Results of the full state tomography at the end of the H-HHL algorithm for different QPUs: IBMQ Burlington, Yorktown (both 5 qubits), Melbourne (15 qubits) and Johannesburg (20 qubits).
        {\bf a.} The Hinton plot of an example of a density matrix, which corresponds to the vector register state -- ancilla and phase registers were filtered out -- that contains the solution of Eq.\,\ref{example}: dashed squares depict the density matrix that corresponds to the ideal solution, solid squares depict the measured density matrix.
        While blue color indicates positive values, red color indicates negative values; the size of a square shows the absolute value.
        {\bf b.} The fidelity $\mathcal{F}_{TOM}$ of the algorithm circuit that was averaged over 70 {\bf NTP} RQCs.
        Since the quantum volume of Burlington, Yorktown and Melbourne processor is the same $V_{\tiny \mbox{Q}}=8$, the fidelity level is similar.
        However, Johannesburg posses slightly higher $V_{\tiny \mbox{Q}}=16$, which is reflected in the fidelity.
    }
    \label{tomography}
\end{figure*} 

The HHL algorithm is improved by using some prior knowledge about the matrix $A$: this classical information allows us to refine the quantum algorithm in order to downsize the noise-sensitive quantum part.
We implement a hybrid HHL algorithm (H-HHL)\,\cite{HybridLloyd} that takes advantage of the fact that some bits of $A^{-1}$ eigenvalues could be the same for any eigenvector.
Since the QPE part encodes eigenvalues into the phase register qubits, we determine the identical bits of eigenvalues and, as a consequence, qubits corresponding to those bits. We treat such phase qubits as classical bits and exclude them from the computational scheme yielding a reduction in the width and in the depth of a circuit.
In general, one can apply an iterative quantum phase estimation algorithm\,\cite{iterativequantumphaseestimation} in order to determine the identical bits of the eigenvalues.

In contract to previous works, here we focus on the large-scale implementation. 
For this purpose, we theoretically introduce and experimentally realize special classes of linear systems that can be solved using the H-HHL algorithm without limitation on the size of the system.\\

\subsection{Benchmarking problem}~~\\
\label{problem}
Here we discuss the class of matrices that we consider in this work.
As an exemplary task, we choose an efficient solution of the following system of linear equations
\begin{equation}
    \begin{cases} 
        \frac{1}{2\pi i}\,\log{\hat{U}}\,\Vec{x}=\Vec{b} \\[2pt] \varrho\left(\frac{1}{2\pi i}\log{\hat{U}}\right) < 1 \\ 
    \end{cases},
    \label{SLE}
\end{equation}
where $\hat{U}$ and $\Vec{b}$ are a given matrix and vector, respectively, and $\varrho(\log{\hat{U}})$ is the spectral radius of $\log{\hat{U}}$. 
Since the logarithmic function is ambiguous, we fix the resulting matrix spectrum such that the largest absolute value of eigenvalues is less than\,1.

To construct the quantum circuit that realize the operator $\hat{U}=e^{iA}$, we 
%do not decompose the exponent matrix into the single-qubit and control NOT gates, but, instead, we
consider the matrix $\hat{U}$ comprising quantum gates.
Furthermore, we choose the computational basis $\ket{b} = \ket{0}$.
The entire family of the gate-based matrices is expressed as a tensor product of $M$ local operators $\hat{U}_i$ that act only on the $i$-dimension subset of the computational circuit:
\begin{equation}
    \hat{U} = \bigotimes_{i=1}^M\,\hat{U}_i.
    \label{tensor_product}
\end{equation}

To illustrate purposes the quantum algorithm capabilities, we consider three types of such $U$ operators.
First of all, we choose the block structure that consists of 2x2 blocks, 4x4 blocks and $\mbox{dim}(U)$ blocks.
We divided the physical operators $\hat{U}$ that correspond to this structure and are commonly implemented in quantum circuits into three groups:
\begin{enumerate}
    \item[{\bf TP$_1$}:] $\hat{U}_{\bf TP_1}$ is the {\bf T}ensor {\bf P}roduct of single-qubit gates $\hat{U}_s$ resulting in dim$(U_i)=2$: this operator does not entangle qubits. See Fig.\,\ref{quantum_circuit}d.
    \item[{\bf TP$_2$}:] $\hat{U}_{\bf TP_2}$ is the {\bf T}ensor {\bf P}roduct of two-qubit operators, dim$(U_i)=2^2$
    that leads to the emergence of two-qubit clusters within which qubits are entangled. See Fig.\,\ref{quantum_circuit}e.
    \item[{\bf NTP}:] $\hat{U}_{\bf NTP}$ is {\bf N}ot a {\bf T}ensor {\bf P}roduct of single- or two-qubit gates that leads to dim$(U_i)=$ dim$(A)$: this operator entangles all qubits. See Fig.\,\ref{quantum_circuit}f.
\end{enumerate}
%
%The quantum circuits for each type are depicted in Fig.\,\ref{quantum_circuit}d,e,f.

The QPE protocol, which is a vital subroutine of the studied algorithm, involves controlled unitary operations and, therefore, is the most complicated part.
Indeed, the realization of an arbitrary control single-qubit gate in $\hat{U}$ requires two CNOTs\,\cite{nielsen}, which leads to a highly complex quantum circuit, e.g. $[\log{N}]$-qubit $\hat{U}_{\bf NTP}$ comprises up to $2\times[\log{N}]$ single-qubit gates.
Thus, we consider the {\it continuous subset} of single-qubit gates $\hat{U}_s$ in a way that the control-$\hat{U}_s$ is implemented via a single CNOT resulting in a dramatic simplification of the algorithm circuit.
The circuit for the control-$\hat{U}_s$ is depicted in Fig.\,\ref{quantum_circuit}g.
%{\color{red} WHAT ARE THE MATRICES A THAT GIVE SUCH SIMPLE CIRCUITS? AT LEAST TP1 EXAMPLE}

In order to demonstrate the classical part of the quantum algorithm, we manually control the spectrum $\{\lambda\}$ of $A$.
For that purpose, we introduce a correcting single-qubit gate $\hat{U}_c$ in the quantum circuit; this assumption significantly simplifies the analysis of a quantum solution and, at the same time, does not lead to loss of generality.
%, and, on the other hand, reduce multi-qubit operators $\hat{U}^2, \hat{U}^4, \dots$ to a single qubit rotation that additionally greatly simplifies the QPE.
%As a result, this choice of matrices yields the fixed spectrum which is comprised of a fixed number of eigenvalues (for more details see Methods); this assumption significantly simplifies the analysis of a quantum solution and, at the same time, does not lead to loss of generality.
Leveraging the full control over the matrix spectrum via the correcting gate $\hat{U}_c$, we immediately eliminate the unnecessary phase qubits without the iterative phase estimation algorithm (for more details see Methods).

As an illustrative example, we adjust the correcting gate in such a way that $\lambda \in \left\{ \frac{1}{8}, \frac{3}{8}, \frac{5}{8}, \frac{7}{8} \right\}$.
As a result $p=3$ phase qubits are required to encode the whole spectrum of $N\times N$ $\hat{U}_{\bf TP_1}$, $\hat{U}_{\bf TP_2}$ or $\hat{U}_{\bf NTP}$ matrices.
The hybrid part allows us to utilize only $p=2$ instead of $3$ phase qubits in order to solve the linear system resulting in $n=[\log{N}]+3$ qubits in total.
The scheme of the H-HHL algorithm is shown in Fig.\,\ref{quantum_circuit}b.
Such a technique allows us to significantly reduce the depth of the circuit, by an order of magnitude for the 50-qubit scheme.
The circuit depth of the simplified H-HHL (red) and the original HHL (blue) as function of a problem size is shown in Fig.\ref{quantum_circuit}c for discussed matrix types.\\

\begin{figure*}[t]
    \noindent\centering{
    \includegraphics[width=180mm]{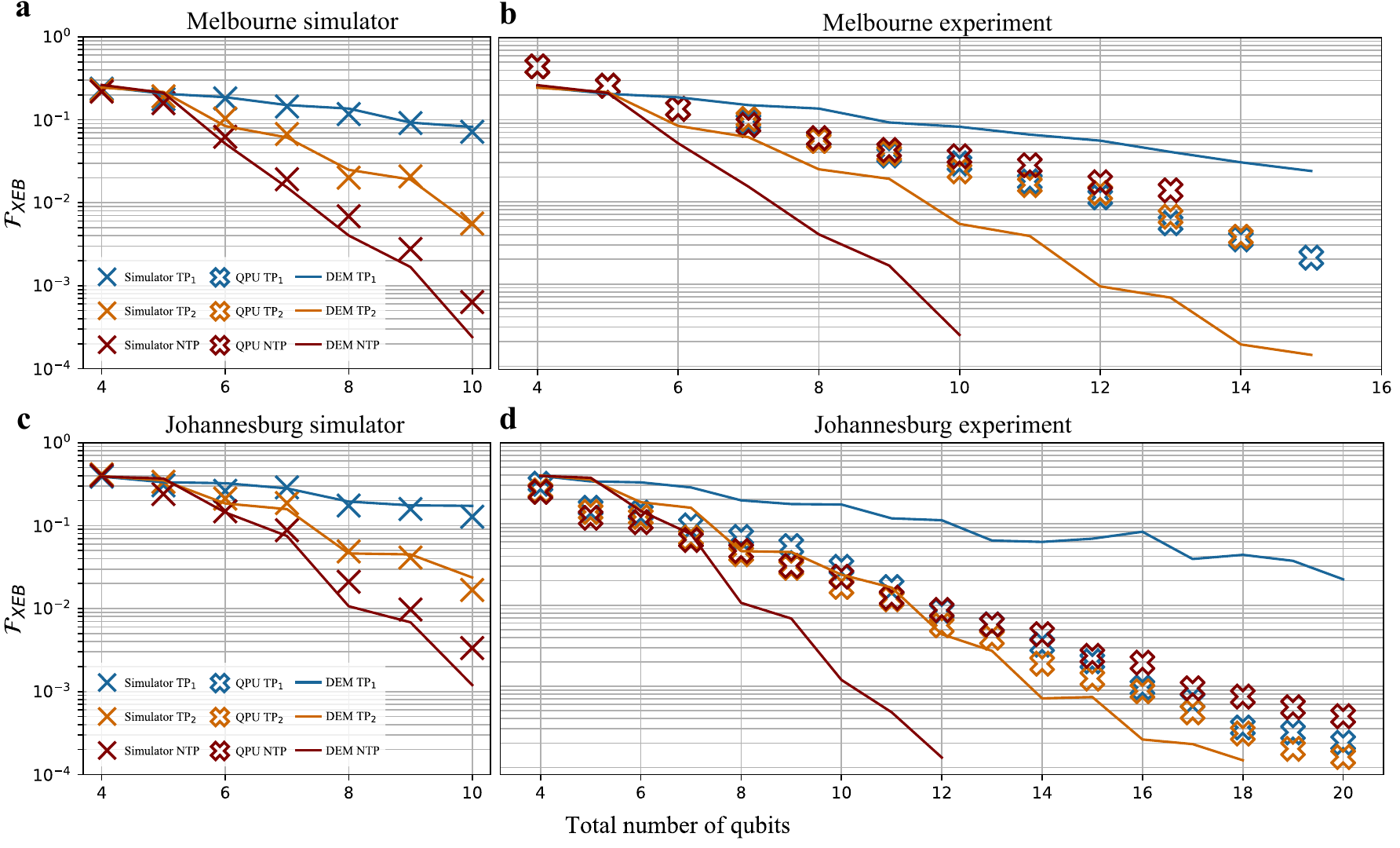}
    }
    \caption{
    Cross entropy fidelity $\mathcal{F}_{XEB}$ of the H-HHL algorithm as a function of a total number of qubits characterized for the IBMQ 15-qubit Melbourne and 20-qubit Johannesburg processors.
    {\bf a.} The fidelity obtained on a simulator with an embedded depolarizing noise model based on the gate and readout errors measured for the IBMQ Melbourne device (filled crosses) for {\bf TP$_1$} (blue), {\bf TP$_2$} (orange) and {\bf NTP} (red) matrices.
    The solid lines indicate the digital error model.
    It is clear that the simulator, which is devoid of any spatial or temporal correlations of gate errors, is described by the DEM.
    {\bf b.} The experimental results measured on the Melbourne QPU (empty crosses); colormap matches the matrix types.
    The same type of measurements were performed for the 20-qubit Johannesburg processor {\bf c.}~on the simulator with a noise model and {\bf d.}~on the real device.
    The experimental results are not fitted by DEM by reason of correlations of control errors, which are specific to IBMQ devices. 
    %\textcolor{orange}{"It would be worth while investigating this further."}
    % Есть обсуждение в основном тексте 
    } \label{xeb}
\end{figure*}

\subsection{Small-scale quantum algorithm implementation}~~\\
Here, we demonstrate the performance of the H-HHL algorithm, which exploits 2 phase and a single ancillary qubit, using real QPUs provided by IBMQ\,\cite{IBMQ}.
%For constructing the quantum circuit, we do not decompose the matrix $e^{iA}$ into the single-qubit and control NOT gates, but, instead, we consider the matrix $\hat{U}$ that is generated by quantum gates assuming that the matrix exponentation is

Primarily, we consider a low number of qubits case where $n \in [4, 7]$, three qubits that are assigned to the phase and ancillary registers.
Under these conditions we employ the full state tomography analysis, which requires $3^{n-3}$ experiments with one circuit; each experiment consists of 8912 runs.
We investigate all matrix types {\bf TP$_1$, TP$_2$}, and {\bf NTP} by analyzing 140 transpiled random quantum circuits (RQCs) for every type for every matrix size.
Let us recall that the randomness is defined by the random single-qubit gates $\hat{U}_s$.
We run the low-width circuits on the IBMQ Burlington (5 qubits), Yorktown (5 qubits), Melbourne (15 qubits) and Johannesburg (20 qubits) QPUs.

First, as an example, let us solve the following system of linear equations
\begin{equation}
    \left[
        \begin{array}{cc}
            \frac{1}{2} & -\frac{3}{8}i \\
            \frac{3}{8}i & \frac{1}{2} 
        \end{array} 
    \right]\,
    \Vec{x}
    =
    \left[
        \begin{array}{c}
            1 \\
            0 
        \end{array}
    \right],
    \label{example}
\end{equation}
where the matrix is determined by the operator $A=\log{\hat{U}_3(\pi/2,0,0)}$ (see Methods).
The normalized solution, in turn, would be encoded into a pure single-qubit quantum state $\ket{x}=\frac{4}{5}\ket{0}+\frac{3}{5}e^{-i\pi/2}\ket{1}$.
The corresponding density matrix $\rho$ is depicted on Fig.\,\ref{tomography}a: dashed squares represent the noiseless solution, solid squares show the solution obtained on the real IBMQ QPUs, blue and red color indicates the positive and negative values, respectively.
High final fidelity ensures that the eigenvector of $\rho$, which corresponds to the maximum eigenvalue, approximates the solution with high precision.
For instance, we find that the density matrix measured on the Johannesburg QPU provides the eigenvector $\ket{\widetilde{x}}=0.81\ket{0}+0.58e^{-i\pi/2.43}\ket{1}$ that corresponds to the maximum eigenvalue, which is the solution with a $|\braket{x|\widetilde{x}}|^2=97.5$\% precision.

Next, we focus on the algorithm performance with only {\bf NTP} matrices, since the fully entangled state is the most interesting in terms of complexity analysis.
The fidelity of the full state tomography averaged over 70 random {\bf NTP} matrices is shown in Fig.\,\ref{tomography}b for different H-HHL circuit widths.
Since the measured quantum volumes\,\cite{Cross2019} $V_{\tiny \mbox{Q}}$ of IBMQ Burlington, Yorktown and Melbourne processors are equal, the fidelity behavior is similar.
However, we find that the $V_{\tiny \mbox{Q}}$ of the Johannesburg QPU is twice as large, resulting in a slightly higher fidelity level.\\

\subsection{Large-scale quantum algorithm implementation}~~\\
Here we consider large circuits performance, but, at first, let us elaborate on a suitable performance metric.
Since the full state tomography requires $3^n$ experiments for $n$-qubit circuit, it is not possible to obtain tomographical fidelity in a reasonable time.
Thus, we employ a similar cross entropy benchmarking approach as in Ref.\,\onlinecite{Supremacy}, which allows us to estimate the algorithm fidelity with only one experiment and single Z-projective measurement that gives us the probability distribution of outcomes.
Let us define the fidelity as follows:
\begin{equation}
    \mathcal{F}_{XEB} = \frac{\sum_{j=1}^M\,(\Vec{p}^{\,e}_j - \Vec{p}^{\,u},~\Vec{p}^{\,t}_j)}{\sum_{j=1}^M\,(\Vec{p}^{\,t}_j - \Vec{p}^{\,u},~\Vec{p}^{\,t}_j)},
    \label{XEB}
\end{equation}
where $(\cdot,\cdot)$ is a scalar product.
The $\Vec{p}^{\,t}_j$ and $\Vec{p}^{\,e}_j$ are probability distribution of outcomes, which correspond to the noiseless implementation and experimental run of $j$th circuit, respectively; $\Vec{p}^{\,u}$ is the uniform probability distribution.
In the above formula, we average the probability distribution over $M$ different $j$th RQCs.
It is clear that the introduced metric shows averaged proximity of the obtained vector projection to the ideal solution rather than to a chaotic state -- such a definition matches the cross entropy fidelity\cite{Supremacy}.
More details on the algorithm characterization can be found in the Supplementary Information (SI).

The fidelity $\mathcal{F}_{XEB}$ of the H-HHL algorithm as a function of the circuit width is presented in Fig.\,\ref{xeb}\,(a,b) for the high-performance simulator with an embedded noise model based on the IBMQ Melbourne device and for the real experiment conducted on Melbourne QPU, respectively; the same data, which was measured on the Johannesburg device, is presented in Fig.\,\ref{xeb}\,(c,d).
The noise model includes the gate and readout errors that were measured beforehand.
Each point is the fidelity that is averaged over 140 RQCs, which were topologically optimized in order to get the minimal depth, and each RQC experiment consists of $10^5$ runs on the simulator as well as on the QPU.

Besides, we consider the digital error model\,\cite{Supremacy} (DEM), which takes into account gates and readout errors and characterizes circuit performance by a set of localized Pauli errors; measured errors are presented in SI. For the estimation of the algorithm performance on a large QPUs via DEM, we assume that only the fidelity of readout $\mathcal{F}_r$, single-qubit gates $\mathcal{F}_{1QG}$ and two-qubit gates $\mathcal{F}_{2QG}$ contribute to the final fidelity resulting in
\begin{equation}
    \mathcal{F}_{XEB} = \mathcal{F}_{r}\cdot\mathcal{F}_{1QE}\cdot\mathcal{F}_{2QE}.
\end{equation}

The digital error approach can be used to predict the fidelity behavior when there is no space or time correlation between gate errors, which is the case in the simulator with an embedded depolarizing noise model, see Fig.\,\ref{xeb}(a,c).
However, for the experiment conducted on IBMQ QPUs, we find that the DEM fails in predictions of the algorithm performance. 
It becomes clear from Fig.\,\ref{xeb}(b,d) that the measured fidelity is almost independent on the matrix type despite the significant changes in corresponding circuits.
Indeed, it was confirmed earlier that gate errors are correlated in some of the IBM superconducting processors\cite{morris2019nonmarkovian}.
Nonetheless, we expect that an advanced equipment combined with the necessary precautions against the correlated errors, e.g. Purcell filters, will provide the fidelity behaviour according to the digital error model as was shown in Ref.\,\onlinecite{Supremacy}.

As a result, in this work we experimentally find the quantum state projection that corresponds to the solution of a $2^{17}\times 2^{17}$ system of linear equations, which appears to be a record in matrix inversion on a gate-based QPU. 
Previously, the $8\times8$ problem was solved with an adiabatic-inspired quantum algorithm\,\cite{Suba2019} on a four-qubit nuclear magnetic resonance device\,\cite{Wen2019}, and the $32\times32$ problem was tackled with variational quantum linear solver\cite{Coles2019} using Rigetti’s 16-qubit superconducting chip.\\

% \subsection{Quantum algorithm implementation, tensor-networks}~~\\
% Here, we present the H-HHL algorithm, which exploits 2 phase and a single ancillary qubit, implemented as a tensor network. 
% {\color{red} !!!WRITE!!!}\\

% \subsection{Quantum algorithm for a local Hamiltonian}~~\\
% {\color{red} !!!WRITE!!!}\\

\section*{Discussion}

\subsection{Classical solution}~~\\
Here we discuss classical algorithms for solving the linear system $A\vec{x}=\vec{b}$, where $A$ has a fixed spectrum and is sparse, which is the case of Eq.\,\ref{SLE}. Our analysis is carried out on {\bf NTP}-type of system, as the most interesting in terms of complexity.

The classical computational cost of solving the system with $N\times N$ matrix using advanced numerical methods is no less than $O(N^{2})$ regardless of mentioned assumptions. 
%The analysis will be carried out only for the most interesting system of the NTP-type. 
%The classical computational cost of solving Eq.\,\ref{SLE} with $N\times N$ matrix, which consists of the logarithm calculation and inversion procedure, by using advanced numerical methods is no less than $O(N^{2})$ regardless of matrix sparsity.
%Let us show that, unfortunately for classical algorithms, fixed spectrum  and matrix decomposition form given by Eq.\,\ref{tensor_product} are unable to simplify the classical solution.
The fixed spectrum can be exploited in the iteration eigenvalue algorithms (IEA)\,\cite{IterationEigenvalueMethods}.
However, such an algorithm involves multiplications of matrices that lead to polynomial scaling in time and memory, since the product of sparse matrices is not guaranteed to be sparse\,\cite{Yuster2004}. Then the IEA cannot be performed in less than $O(N^2)$ operations. 
On the other hand, in order to compute $\log{\hat{U}}$ one can utilize matrix diagonalization\cite{Loring2014} or the inverse scaling and squaring method\,\cite{LogMatrix}.
Such techniques require at least an exponential number of operations or addresses to matrix elements.
If the sparse $\log{\hat{U}}$ is obtained, one may use iterative algorithms to solve the linear system afterward, e.g. by applying Kaczmars method\cite{Strohmer2008}, which has also the $O(N^2)$ scaling. \\

\subsection{Quantum algorithmic advantage}~~\\
\begin{figure}[t]
    \noindent\centering{
    \includegraphics[width=85mm]{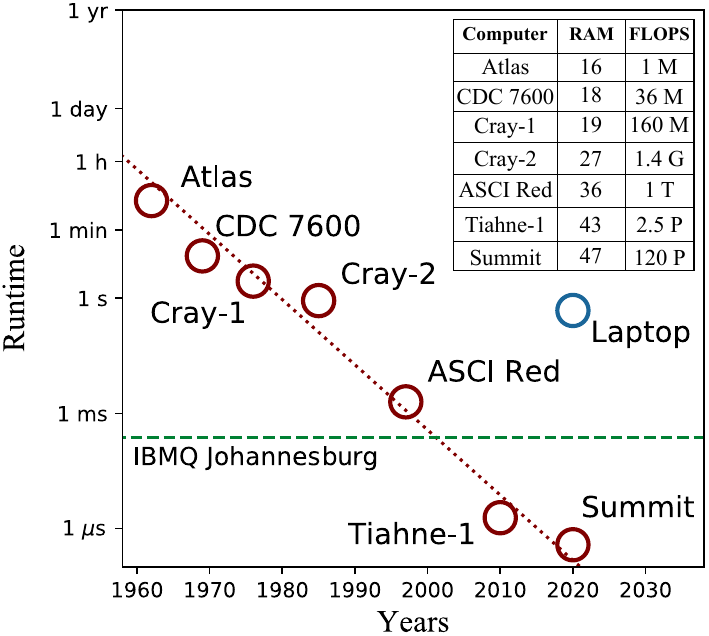}
    }
    \caption{
    Runtime of the equal-fidelity classical solution of a $2^{17}\times 2^{17}$ linear system with {\bf NTP} matrix on historical supercomputers.
    The fidelity level of a IBMQ Johannesburg processor, which solves the problem in 200\,$\mu$s (dashed green line), is 0.04\%.
    The runtime was calculated by estimating the whole 20-qubit circuit simulation runtime and multiplying it on fidelity.
    In case when the supercomputer's memory is not enough we divide the circuit in two and simulate all paths using Schr\"{o}dinger-Feynman approach (see Methods).
    The Atlas supercomputer (1962) would solve the system with the same fidelity in 17\,minutes, CDC 7600 (1969) in 30\,seconds, Cray-1 (1976) and Cray-2 (1985) in 6.3\,seconds and 0.9\,seconds, respectively, Intel ASCI Red (1997) in 1.2\,ms.
    While XXth century devices are inferior to the 20-qubit quantum computer, present supercomputers Tiahne-1 (2010) and Summit (2020) surpass the IBMQ QPU.
    Nonetheless, regular modern laptops (10\,GFLOPS) are less efficient than the quantum processor.
    Runtime fit (dotted red line) indicates the Moore's law.
    The embedded table shows RAM in the number of qubits, which can be fitted into memory, and performance in FLOPS.
    } \label{old_computers}
\end{figure}
As we have shown before, while the runtime of the classical algorithms that solves $N$x$N$ linear system is $O(N^2)$, the runtime of the hybrid quantum algorithm has $O(\log{N})$ scaling.
We expect that future improvements in classical algorithms and hardware will provide a considerable reduction in runtime and computational resources, however, persistent enhancement of quantum software allows hybrid quantum approaches to consistently outperform classical solutions.
Besides, we find that the high-performance circuit emulation\cite{Supremacy} is considerably faster than the direct solution: O($N^2$) vs O$(N\,\log{N})$, indicating that simulation of a quantum circuit is essentially an efficient tool for the solution and a proper reference for the quantum hardware to compare.
Here, we consider the straightforward Schr\"{o}dinger-Feynman simulation presented in Ref.\,\onlinecite{Supremacy}.

For algorithmic benchmarking, let us compare the efficiency of the $2^{17}\times2^{17}$ {\bf NTP} problem solution implemented on a QPU and a Schr\"{o}dinger-Feynman simulation implemented on the most significant supercomputers in history.
In order to properly compare classical approaches and quantum devices, we must take into account the final fidelity,  since the equal-fidelity computation time scales linearly with $\mathcal{F}_{XEB}$ as was shown in \onlinecite{Supremacy}.
While the net calculation time on a quantum device with 0.04\% fidelity does not exceed 1\,ms (assuming several runs to ensure the ancillary qubit is measured in $\ket{1}$), the equal-fidelity classical solution would require 7 orders of magnitude higher time for the supercomputers from 1960, 5 orders for devices from 1970 and few times higher for computers from 1990; the detailed comparison is shown in Fig.\,\ref{old_computers}.
Note that the estimations for the simulation performance are over evaluated for supercomputers that do not have enough memory to process the whole circuit. We thus omit the final vector recreation runtime since the estimation of such an operation is a highly complicated task.
Such a dramatic increase in speed is an experimental evidence that the hybrid quantum computing will surpass the classical computing industry that reigned in the XXth century.

We want to emphasize that our comparison was made with the most straightforward simulations that include matrix multiplications.
However, we expect that more advanced quantum simulation approaches, such as tensor networks\cite{Feynman, Zhou2020}, may allow for a solution in poly-logarithmic time.
Nevertheless, this kind of quantum-inspired solution would show the power of quantum algorithms even more vastly.
The realization of these methods is a subject of the forthcoming works.\\

\subsection{Near-term applications}~~\\
Finally, we discuss the possible practical applications of the H-HHL algorithm in the framework of the posed problem.
Despite the fact that the fidelity level is low, the quantum algorithm still provides the solution with a tremendously higher probability than a random guess.
% %
% \begin{figure}[t]
%     \noindent\centering{
%     \includegraphics[width=80mm]{Runtimes1.pdf}
%     }
%     \caption{
%     Runtime of the equal-fidelity classical solution (solid lines) on state-of-the-art 100K cores supercomputer compared to the hybrid quantum solution (dashed line) processing large quantum circuit; colormap matches matrix types.
%     The fidelity level is estimated based on the digital error model involving gate and readout errors from Ref.\,\onlinecite{Supremacy}.
%     The change in the slope of the classical runtime curve is caused by a change in a circuit emulation type: we assume that the 47 qubit circuit is the largest system that can be fitted into supercomputer's RAM, therefore we must split the large circuit into parts and simulate each part in series.
%     We expect that the quantum solution provides an advantage over the classical one when the {\bf TP$_1$}, {\bf TP$_2$} and {\bf NTP} problem size is larger than 44, 51 and 55 qubits, respectively.
%     } \label{supremacy}
% \end{figure}
% %

The exact realization of our algorithm allows for addressing some tangible problems, for instance, in the context of Markov processes, one could obtain the generator of a known stochastic $P$ matrix -- a transition rate matrix $Q = \log{P}$\cite{higham_2008}.
Then, we can solve the linear equation $Q\Vec{f}\cdot\Delta t = \Delta\Vec{f}$ at some point of the process. 
Here, $\Vec{f}$ is the distribution of state probabilities and $\Delta \Vec{f}/\Delta t$ is a probability current, which is supposed to be known. 
Upon obtaining $\ket{f}$ and measuring $\bra{f}M\ket{f}$ we check that the probability of a specific state is non-zero.

An analogous problem can be addressed in the control theory.
Let us suppose that a discrete-time process with time step $\Delta t$: $\Vec{x}_{n+1} = A\Vec{x}_n$ is modelled by a continuous one: $\dot{\Vec{x}} = B\Vec{x}$. 
Since $A = \exp(B\Delta t)$, one could obtain the vector $\Vec{x}$ by solving $\dot{\Vec{x}} = B\Vec{x}$ at some point of the process if $\dot{\Vec{x}}$ is provided. 
By averaging proper Hermitian operators $\bra{x}M\ket{x}$ we determine whether some components of $\Vec{x}$ are non-zero.

The algorithm can be modified by applying matrix simulation techniques for $A$ in order to solve $A\Vec{x}=\Vec{b}$ as it was originally proposed in Ref.\,\onlinecite{Lloyd}.
In that case, a wider range of problems can be approached.
For instance, it was proposed to use the SWAP test in order to determine whether solutions of different systems coincide.
Such linear systems are also used to solve partial differential equations, e.g. one can find the electromagnetic field energy in some region using the Poisson equation. 
One of the promising applications related to deep neural network training was discussed in Ref.\,\onlinecite{Zhao2019}: since the extension of the Bayesian approach to deep architectures is a serious challenge, one can exploit the hybrid quantum HHL algorithm developed for Gaussian processes in order calculate a model's predictor.

\bigskip

\section*{Conclusion}~~\\

We implemented the quantum hybrid HHL algorithm solving a system of linear equation by fast matrix inversion.
The matrix, in turn, is approximated by a unitary transformation, which was dictated by the sequences of single-qubit rotations and CNOT gates.
The size of the linear system grows exponentially with the increasing number of qubits. 
Using the state-of-art 50 qubits processor, one can invert the $10^{14}\times10^{14}$ matrix, which is far beyond the capabilities of the present state-of-the-art supercomputers.

We probed the algorithm on the simulator with an embedded noise model, which is devoid of correlations in gate errors, and on the real IBMQ QPUs with 5, 15 and 20 qubits.
The implemented quantum solution of a $2^{17}$-dimensional problem, which is a record for a linear system solution on quantum computers, demonstrates algorithmic quantum advantage over classical solutions.

% {\color{red} !!!REWRITE!!!
% Furthermore, we estimated the fidelity level of the algorithm implemented on a next-generation 50+ qubit processors and found that the system cannot be solved faster with the same fidelity on a supercomputer neither directly nor by using special techniques such as a high-performance simulation of a quantum circuit.
% Observed exponential scaling in the equal-fidelity classical computation runtime indicates that NISQ devices, which exploit only polynomial time resources, exponentially outperform existing classical analogues.
% \color{red} !!!REWRITE!!!}

Our experimental results and theoretical estimates hold high potential for solving large linear systems utilizing state-of-the-art NISQ computers and quantum algorithms in general. 
We intent to probe the H-HHL algorithm both on the cutting edge low-noise QPUs and on high-performance classical simulator machines.
%We intend to probe the H-HHL algorithm on the cutting edge low-noise QPUs and collect experimental data in order to give a chance to the future computational devices to verify the fidelity.
We envision that the planned experiments will stimulate major players in the quantum computing industry to demonstrate the hardware and software achieving quantum advantage.

\bigskip

\small{

\begin{methods}~~~\\

\subsection{Linear system construction}~~\\
Here, we consider the implementation of {\bf TP$_1$}, {\bf TP$_2$} and {\bf NTP} matrix types in more detail.
In order to construct quantum circuits we use Qiskit open-source framework\cite{Qiskit} that allows us to generate $\hat{U}$ from single-qubit and two-qubit CNOT gates.
An arbitrary single-qubit gate is decomposed by Qiskit into physical gates: the $\hat{U}_3$ operator, which is the most general gate, defines as follows:
\begin{equation}
    \hat{U}_3(\theta, \phi, \lambda) = 
    \begin{pmatrix}
        \cos\frac{\theta}{2} & -e^{i\lambda} \sin\frac{\theta}{2} \\
        e^{i \phi} \sin\frac{\theta}{2} & e^{i\lambda + i\phi} \cos\frac{\theta}{2}
    \end{pmatrix}.
    \label{u3}
\end{equation}

Since the QPE protocol, which is a vital part of the H-HHL algorithm, involves controlled unitary operations, we need to extend single-qubit gates $\hat{U}_s$ and CNOTs in $\hat{U}$ to a control $\hat{CU}_s$ and Toffoli gates, respectively\cite{nielsen}.
However, realization of an arbitrary control single-qubit gate $\hat{CU}_s$ requires two CNOTs that leads to highly complex quantum circuit.
Thus, we consider subclass of a single-qubit $\hat{U}_s$ that the $\hat{CU}_s$ requires only one CNOT resulting in dramatic simplification of the H-HHL circuit.

Let us define operator $\hat{U}_s=\hat{U}_3 \{\hat{X}, \hat{Y}, \hat{Z}\} \hat{U}_3^{\dagger}$, where any of the Pauli matrices $\{\hat{X}, \hat{Y}, \hat{Z}\}$ can be chosen.
Such a form leads to the following C$\hat{U}_s$ representation:
\begin{equation}
    \begin{aligned}
        C\hat{U}_s &= \ket{0} \bra{0} \otimes \hat{U}_3 \hat{U}_3^{\dagger} + \ket{1} \bra{1} \otimes \hat{U}_3 \{\hat{X}, \hat{Y}, \hat{Z}\} \hat{U}_3^{\dagger} = \\
        &= \ket{0} \bra{0} \otimes \hat{I} + \ket{1} \bra{1} \otimes \hat{U}_s,
    \end{aligned}
\end{equation}
where the Pauli gate is applied only when the control qubit is in the excited state, see Fig.\,\ref{quantum_circuit}g.
It is clear that $\hat{U}_s$ gate is essentially an arbitrary Hermitian $\hat{U}_3$ gate, which can be build according to definition Eq.\,\ref{u3} with following condition:
\begin{equation}
    \lambda + \phi = (2 k - 1) \pi \text{, where } k \in \mathbb{Z}.
\end{equation}

Such a definition ensures that all controlled single-qubit gates contain only one two-qubit operation and, since $\hat{U}_s$ is a Hermitian gate, it has only two eigenvalues $\{1, -1\}$.
Unfortunately, it leads to another problem: studied $\log(\hat{U}_{\bf TP_1})$, $\log(\hat{U}_{\bf TP_2})$, $\log(\hat{U}_{\bf NTP})$ matrices have zero eigenvalues, which is unacceptable for the HHL algorithm\cite{Lloyd}.
In order to avoid such an issue we add the following correcting gate to the first register of $\hat{U}$:
\begin{equation}
    \hat{U}_3(\pi/2, 0, 0) = 
    \frac{1}{\sqrt{2}}
    \begin{pmatrix}
        1 &  -1 \\
        1 &  1
    \end{pmatrix}.
\end{equation}
The introduction of a such correcting gate leads to two important and fruitful features: (i) the spectrum of all matrix types is exp[$2\pi i\cdot\{\frac{1}{8}, \frac{3}{8}, \frac{5}{8}, \frac{7}{8}\}$]; 
(ii) $\hat{U}^2_{\bf TP_1}=\hat{U}^2_{\bf TP_2}=\hat{U}^2_{\bf NTP}=\hat{U}_3(\pi,0,0)\otimes\hat{I}$ that greatly simplifies the QPE part.

We exploit the knowledge about matrix spectrum in the classical part of the H-HHL algorithm.
Since the binary form of $\frac{1}{2\pi i}\log(\hat{U})$ eigenvalues contain three bits, one needs three phase qubits to store whole spectrum.
It is easy to show that the last bit of all eigenvalue is 1 if we write down four eigenvalues in binary representation: $\lambda = \frac{\{0,1\}}{2} + \frac{\{0,1\}}{4} + \frac{1}{8}$, where the different combinations of $\{0,1\}$ corresponds to different eigenvalues. 
Therefore, we immediately eliminate one phase qubit without conducting the iterative phase estimation procedure resulting in $p=2$.
Hence, we can utilize $n=[\log{N}]+3$~qubits overall for each matrix type.
The 5-qubit examples of $\hat{U}_{\bf TP_1}$, $\hat{U}_{\bf TP_2}$ and $\hat{U}_{\bf NTP}$ operators are depicted on Fig.\,\ref{quantum_circuit}d, Fig.\,\ref{quantum_circuit}e and Fig.\,\ref{quantum_circuit}f, respectively.

\bigskip

\subsection{Circuit emulation}~~\\
Since the classical computational cost of solving Eq.\,\ref{SLE} with $N\times N$ matrix is no less than $O(N^{2})$,
%Unfortunately for classical algorithms, fixed spectrum and matrix decomposition given by Eq.\,\ref{tensor_product} are unable to simplify the classical solution; for more detailed analyses see SI.
an alternative method, the high-performance simulation of the H-HHL quantum circuit, should be considered.
The simulation that consists of Schr\"{o}dinger algorithm (SA)\cite{Supremacy}, on one hand, is straightforward and provides a great speed in processing low-width circuits in comparison to the other simulations.
On the other hand, the Schr\"{o}dinger algorithm requires a significant amount of RAM when processing many-qubits circuits.
Such a requirement is difficult to meet since the memory of modern supercomputers is substantially limited.
However, the Schr\"{o}dinger-Feynman algorithm\cite{Supremacy} (SFA) solves the memory issues inherent in a SA, which presumably makes it the fastest simulation of high-deep high-width quantum circuits.

Primarily, in order to estimate the classical simulation runtime, we perform Schr\"{o}dinger simulations, which are essential building blocks of SFA; the runtime scaling of a single $n$-qubit SA is $T_{SA} = C \cdot n \cdot 2^n$.
We exploit a POWER8 processor with 160 cores and 512\,Gb of RAM\cite{Qiskit} in order to estimate the time constant $C$ -- evaluating the runtime for each circuit width we fit the dependence and find that $C_{\bf TP_1} = 5 \cdot 10^{-9}$\,s, $C_{\bf TP_2} \approx 14 \cdot 10^{-9}$\,s and $C_{\bf NTP} \approx 17 \cdot 10^{-9}$\,s, which we assume are scaled linearly in number of cores.
Since modern supercomputers have roughly $100$K cores, we expect the constants to be $C_{\bf TP_1} \approx 10^{-12}$\,s, $C_{\bf TP_2} \approx 2.8\cdot10^{-12}$\,s and $C_{\bf NTP} \approx 3.4\cdot10^{-12}$\,s in the best scenario.

For memory estimates, while the state-of-the-art supercomputers have 3\,PB of RAM, we suppose that one can store a $2^{47}$-dimensional vector using 8 bytes to encode single complex number.
Hence, SFA allows us to split the quantum circuit in the ratio $(n - \widetilde{n}):\widetilde{n}$, where $\widetilde{n} < n - \widetilde{n} \leq 47$. 
The execution time of the Schr\"{o}dinger-Feynman method is $T_{SFA} = C\,(n-\widetilde{n})\cdot2^{n-\widetilde{n}} \cdot N_p$, where $N_p$ is the number of paths in final state vector calculation\cite{Supremacy}. 
The scaling constant $C$ is the same as in the Schr\"{o}dinger algorithm, since we are forced to use almost all RAM and algorithm parallelization becomes an elusive task.
%For instance, we expect that the classical solution with $2^{50}\times2^{50}$ matrices will be evaluated in $T_{SFA}(\hat{U}_{\bf TP_1}) \approx 10$ months, $T_{SFA}(\hat{U}_{\bf TP_2}) \approx 5$ and $T_{SFA}(\hat{U}_{\bf NTP}) \approx 400$ million years of classical computations, respectively.

In order to compare 20-qubit QPUs with classical supercomputers, we recalculate constants $C$ in seconds per Flops; for the {\bf NTP} circuit we found that $T_{SA}~=~1.5\cdot10^5\cdot\,n\cdot2^n$/Flops\,s.
For the characterization of the equal-fidelity calculation, we 
assume that the simulation runtime scales linear with fidelity, therefore the equal-fidelity runtime is $\mathcal{F}_{XEB}\,T_{SA}$.
However, when it is impossible to store the whole vector in the memory we divide our circuit into $16:4$ ratio.
In that case, we obtain a single Toffoli gate on the cut for the single QPE step, which have two control qubits above the cut and controlled qubit bellow the cut. 
This gate should be decomposed in order to simulate two parts independently and combined afterwards to get a solution.
There are $2^2$ paths for such Toffoli gate that gives us $N_p = 2^4$ in total (due to QPE and inverse QPE).
%cut the circuit in the same way as the SFA, which generates $2^k$ sub-circuits, and run only $\mathcal{F}_{XEB}\cdot2^k$ sub-circuits via SA, e.g. for the {\bf NTP} case, we find that the 20-qubit circuit should be cut into $18:2$ ratio; a single simulation of each sub-circuit is equivalent to the solution with 0.04\% fidelity.
Using the calculated runtime for $2^4$ $16$-qubit circuit emulations, we estimate the equal-fidelity runtime for the supercomputers and plot the data in Fig.\,\ref{old_computers}.

\bigskip

\end{methods}
}

\bigskip
\bibliographystyle{naturemag}
\bibliography{Master}

\hskip 1pt

\begin{addendum}

\item [Acknowledgements]%\section{Acknowledgements}
This work was supported by the Government of the Russian Federation (Agreement 05.Y09.21.0018),  
the Russian Foundation for Basic Research under Grants No. 18-02-00642A, and 
the Foundation for the Advancement of Theoretical Physics and Mathematics "BASIS".
The work of A.G. at Argonne was supported by the U.S. Department of Energy, Office of Science, Basic Energy Sciences, Materials Sciences and Engineering Division. 
The research used resources of the Oak Ridge Leadership Computing Facility, which is a DOE Office of Science User Facility supported under Contract DE-AC05-00OR22725.
G.S.P. acknowledges support from the Academy of Finland through the Finnish Center of Excellence in Quantum Technology QTF (project 312296). The work of V.M.V., M.R.P. A.I.P., A.A.M., and G.B.L. was supported by Terra Quantum AG.

\bigskip

\item [Author contribution]
A.I.P., M.R.P. and A.A.M. devised the algorithm.
A.I.P., M.R.P. and A.G. wrote the code and performed the experiment; 
%the data was processed by M.R.P., A.A.M., A.I.P. and A.A.N.
the algorithm performance was analyzed by M.R.P., A.A.M., A.I.P. and A.A.N.
M.R.P., A.I.P., A.A.M. and V.M.V. wrote the manuscript with inputs from all authors.
G.S.P., V.M.V. and G.B.L. supervised the project.
All authors discussed the results and contributed to the work.

\item[Competing interests] The authors declare that they have no competing
financial interests.

\item[Correspondence] Correspondence and requests for materials should be
addressed to V.M.V. (vv@terraquantum.swiss).

\clearpage

\section*{Supplementary Information}~~\\
\setcounter{figure}{0} 
\setcounter{equation}{0}
\renewcommand{\theequation}{S\arabic{equation}}
\renewcommand{\thefigure}{S\arabic{figure}}
\bigskip

\renewcommand{\thetable}{S\arabic{table}}

\subsection{Problem-solving circuit complexity}~~\\
In order to estimate the potential size of the quantum circuit, we employ the best existing superconducting quantum processors, such as IBMQ Melbourne (15 qubits), Johannesburg (20 qubits), Rochester (53 qubits) and Google Sycamore (53 qubits).
We use the QPUs coupling map and evaluate the circuit depth -- a leading characteristic of the circuit complexity including the amount of a single- and two-qubit gates.

Since the circuit of the quantum algorithm is usually elaborated for the all-to-all connectivity, we can not instantly run it on a real QPU, where the connectivity is limited.
Firstly, one needs to transform the quantum circuit to fit the device topology -- we will refer to this operation as transpiling.
In order to achieve higher final fidelity one needs to solve the transpiling problem of finding the optimal transformation that maximizes the final fidelity.
Unfortunately, finding the best transformation is an NP-hard problem, therefore we use a brute force search over dozens of transpile options and pick the best circuit decomposition.

The dependence of the circuit depth on the circuit width (number of qubits) used in the H-HHL algorithm is indicated in Fig.\,\ref{CNOT}a for medium-size IBMQ Melbourne and Johannesburg QPUs, and on Fig.\,\ref{CNOT}b for the largest modern Sycamore and Rochester QPUs for all circuit types.
In order to obtain the circuit depth, we averaged over 140 transpiled random quantum circuits (RQCs) that realize $\hat{U}$ for each matrix type and size.
\begin{figure}[h]
    \noindent\centering{
    \includegraphics[width=86mm]{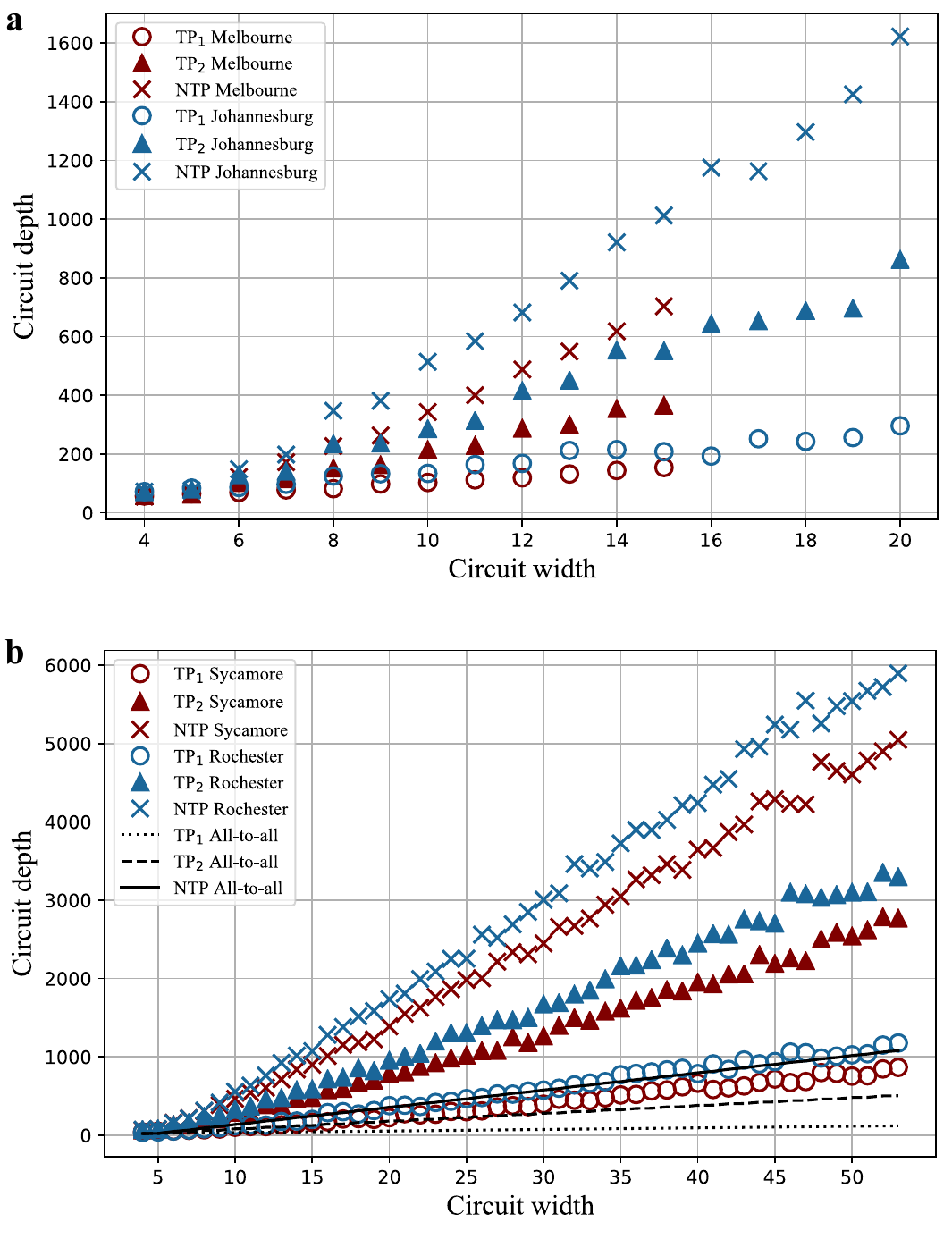}
    }
    \caption{
    The depth of transpiled H-HHL circuit as a function of the circuit width, which corresponds to the {\bf TP$_1$} (circles), {\bf TP$_2$} (triangles) and  {\bf NTP} (crosses) matrix processing.
    {\bf a.} Depth of the circuit implemented on medium-size IBMQ 15 qubit Melbourne (red) and 20 qubit Johannesburg (blue) QPUs. 
    {\bf b.} Depth of the circuit implemented on large-size IBMQ Rochester (blue) and Google Sycamore (red) processors: both 53 qubits.
    While the 53-qubit circuit implemented on the all-to-all coupling map consists of $\leq10^3$ two-qubit gates (solid, dashed and dotted black lines), the circuit implemented on the IBMQ Rochester device requires $\geq10^3$ entangling gates for {\bf TP$_1$}, $\sim 3\cdot10^3$ for {\bf TP$_2$} and $\sim 6\cdot10^3$ for {\bf NTP} matrices.
    Google Sycamore's topology allows for a 26\% reduction in circuit depth compared to the Rochester QPU, providing a significant improvement in fidelity.
    } \label{CNOT}
\end{figure}

It is clear that the poor connectivity has a huge impact on the circuit depth as well as on the expected final. 
Since the 53-qubit circuit implemented on the all-to-all coupling map consists of $\sim[10^2, 5\cdot10^2, 10^3]$ two-qubit gates for {\bf TP$_1$}, {\bf TP$_2$} and {\bf NTP} type, respectively, the same circuit implemented on the IBMQ Rochester device requires $\sim[10^3, 3\cdot10^3, 6\cdot10^3]$ CNOTs.
Google Sycamore's topology allows for a 26\% reduction in CNOTs compared to the Rochester QPU resulting in a significant improvement in fidelity.

For illustrative purposes, we consider a quantum volume $V_{\tiny \mbox{Q}}$ of QPUs, a single-number metric presented by IBM that can be experimentally obtained using random quantum circuits\cite{Cross2019}.
A quantum volume quantifies the largest RQCs of equal width and depth that the computer successfully implements.
We measure the quantum volume of the IBM processors and estimate such a metric for the Sycamore machine.
We find that all 5-qubit IBMQ devices, 15-qubit Melbourne, and 53-qubit Rochester processors, have $V_{\tiny \mbox{Q}}=8$, the 20-qubit Johannesburg has $V_{\tiny \mbox{Q}}=16$ and the Google Sycamore processor has $V_{\tiny \mbox{Q}}=32$.
Thus, it is not a surprise that much lesser amount of two-qubit interactions is required for the Sycamore than for the Rochester.\\

\bigskip

\bigskip

\bigskip

\subsection{Large-scale algorithm characterization}~~\\
The XEB fidelity defined in Eq.\,\ref{XEB} originates from the depolarizing channel model, which was used in order to describe the error propagation process.
The introduced metric shows averaged proximity of the experimentally obtained outcomes distribution $p^e$ to the ideal outcomes distribution $p^t$ rather than to a uniform distribution $p^u=\{1/2^n, 1/2^n, \cdots, 1/2^n\}$, which originates from the chaotic quantum state $\hat{\rho}=\hat{I}/2^n$.
In Eq.\,\ref{XEB} the number $(p^e, p^t)-(p^u, p^t)$ in the numerator shows the proximity that we are interested in, and the denominator is a normalization coefficient.

However, let us discuss an alternative method for algorithm characterization, which also requires just $O(n)$ experiments, in the absence of gate error correlations.
In essence, our quantum algorithm realizes the unitary transformation $\hat{U}_{HHL}$ applied to the $n$-qubit ground state $\ket{0}$.
One can construct and run the $\hat{U}_{HHL}\hat{U}^\dagger_{HHL}$ circuit and compare the outcomes with the ground state projection by using the Eq.\,\ref{XEB} -- we can estimate the fidelity of the quantum algorithm as a square root of an obtained value.
Such a method is applicable in the supremacy regime, however, it requires a considerable low total noise level that ensures high final fidelity.
\begin{figure}[h]
    \noindent\centering{
    \includegraphics[width=80mm]{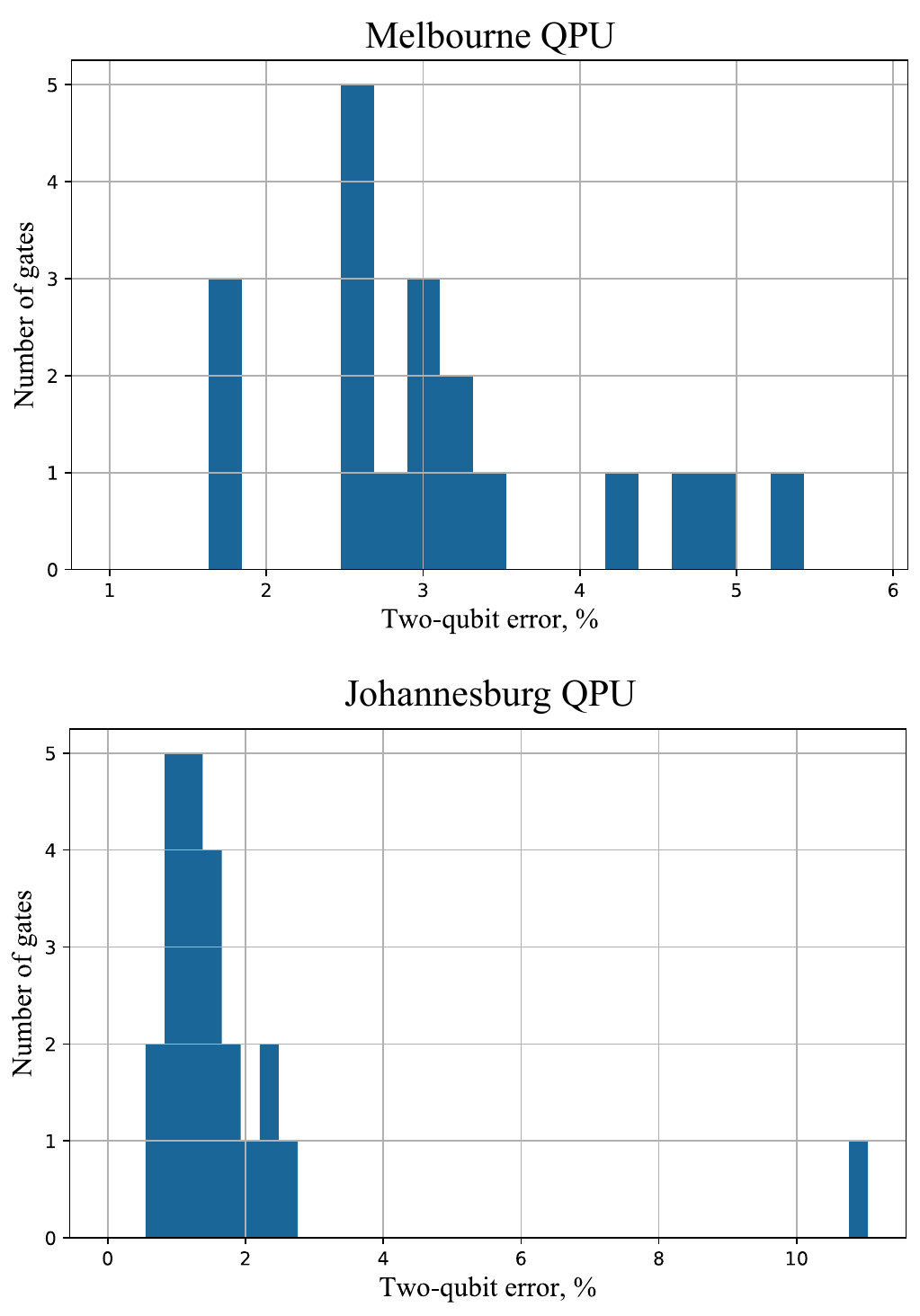}
    }
    \caption{ 
    Measured CNOTs error distribution on a Melbourne and Johannesburg IBMQ QPUs.
    It is clear that the Johannesburg device is less noisy than a Melbourne, which is reflected in the  quantum volume: $V_{\tiny \mbox{Q}}=16$ for Johannesburg and $V_{\tiny \mbox{Q}}=8$ for Melbourne.
    } \label{errors}
\end{figure}

\bigskip

\subsection{Digital error model}~~\\
The digital error model characterizes the circuit performance by a set of localized Pauli error.
The probability of a particular Pauli gate appearing is defined by the control gate error, which was measured for a 15-qubit Melbourne and 20-qubit Johannesburg devices.
The error of a two-qubit gate is an order of a magnitude higher than for a single-qubit error; we plot the CNOTs error distribution for a Melbourne and Johannesburg QPU in Fig.\,\ref{errors}.

\end{addendum}

\end{document}